\documentclass[12pt,]{article}
\usepackage{lmodern}
\usepackage{amssymb,amsmath}
\usepackage{ifxetex,ifluatex}
\usepackage{fixltx2e} 
\ifnum 0\ifxetex 1\fi\ifluatex 1\fi=0 
  \usepackage[T1]{fontenc}
  \usepackage[utf8]{inputenc}
\else 
  \ifxetex
    \usepackage{mathspec}
  \else
    \usepackage{fontspec}
  \fi
  \defaultfontfeatures{Ligatures=TeX,Scale=MatchLowercase}
\fi
\IfFileExists{upquote.sty}{\usepackage{upquote}}{}
\IfFileExists{microtype.sty}{%
\usepackage{microtype}
\UseMicrotypeSet[protrusion]{basicmath} 
}{}
\usepackage[margin=1in]{geometry}
\usepackage{hyperref}
\hypersetup{unicode=true,
            pdftitle={Forecasting Fertility using Parametric Mixture Models},
            pdfauthor={Jason Hilton\^{}1; Erengul Dodd\^{}1; Jonathan J. Forster\^{}2; Peter W.F Smith\^{}1; Jakub Bijak\^{}1; \^{}1 Centre for Population Change, University of Southampton; \^{}2 Department of Statistics, University of Warwick; Correspondence: J.D.Hilton@soton.ac.uk},
            pdfborder={0 0 0},
            breaklinks=true}
\usepackage{longtable,booktabs}
\usepackage{graphicx,grffile}
\makeatletter
\def\maxwidth{\ifdim\Gin@nat@width>\linewidth\linewidth\else\Gin@nat@width\fi}
\def\maxheight{\ifdim\Gin@nat@height>\textheight\textheight\else\Gin@nat@height\fi}
\makeatother
\setkeys{Gin}{width=\maxwidth,height=\maxheight,keepaspectratio}
\IfFileExists{parskip.sty}{%
\usepackage{parskip}
}{
\setlength{\parindent}{0pt}
\setlength{\parskip}{6pt plus 2pt minus 1pt}
}
\providecommand{\tightlist}{%
  \setlength{\itemsep}{0pt}\setlength{\parskip}{0pt}}
\ifx\paragraph\undefined\else
\let\oldparagraph\paragraph
\renewcommand{\paragraph}[1]{\oldparagraph{#1}\mbox{}}
\fi
\ifx\subparagraph\undefined\else
\let\oldsubparagraph\subparagraph
\renewcommand{\subparagraph}[1]{\oldsubparagraph{#1}\mbox{}}
\fi

\let\rmarkdownfootnote\footnote%
\def\footnote{\protect\rmarkdownfootnote}

\usepackage{titling}

  \title{Forecasting Fertility using Parametric Mixture Models}
    \pretitle{\vspace{\droptitle}\centering\huge}
  \posttitle{\par}
    \author{Jason Hilton\(^1\) \\ Erengul Dodd\(^1\) \\ Jonathan J. Forster\(^2\) \\ Peter W.F Smith\(^1\) \\ Jakub Bijak\(^1\) \\ \(^1\) Centre for Population Change, University of Southampton \\ \(^2\) Department of Statistics, University of Warwick \\ Correspondence:
\href{mailto:J.D.Hilton@soton.ac.uk}{\nolinkurl{J.D.Hilton@soton.ac.uk}} \\ \newpage}
    \preauthor{\centering\large\emph}
  \postauthor{\par}
    \date{}
    \predate{}\postdate{}
  
\usepackage{booktabs}
\usepackage{longtable}
\usepackage{array}
\usepackage{multirow}
\usepackage{wrapfig}
\usepackage{float}
\usepackage{colortbl}
\usepackage{pdflscape}
\usepackage{tabu}
\usepackage{threeparttable}
\usepackage{threeparttablex}
\usepackage[normalem]{ulem}
\usepackage{makecell}
\usepackage{xcolor}

\usepackage{caption}
\begin{document}
\maketitle
\begin{abstract}
This paper sets out a forecasting method that employs a mixture of
parametric functions to capture the pattern of fertility with respect to
age. The overall level of cohort fertility is decomposed over the range
of fertile ages using a mixture of parametric density functions. The
level of fertility and the parameters describing the shape of the
fertility curve are projected foward using time series methods. The
model is estimated within a Bayesian framework, allowing predictive
distributions of future fertility rates to be produced that naturally
incorporate both time series and parametric uncertainty. A number of
choices are possible for the precise form of the functions used in the
two-component mixtures. The performance of several model variants is
tested on data from four countries; England and Wales, the USA, Sweden
and France. The former two countries exhibit multi-modality in their
fertility rate curves as a function of age, while the latter two are
largely uni-modal. The models are estimated using Hamiltonian Monte
Carlo and the \texttt{stan} software package on data covering the period
up to 2006, with the period 2007-2016 held back for assessment purposes.
Forecast performance is found to be comparable to other models
identified as producing accurate fertility forecasts in the literature.
\newpage
\end{abstract}

\section{Introduction}\label{introduction}

Human fertility is a complicated, dynamic process that is influenced by
a wide range of biological, social and economic factors as diverse as
changes in social norms around female education and employment; the
availability and cost of childcare; and wider trends in wages,
disposable income and employment (see Balbo, Billari, and Mills
\protect\hyperlink{ref-Balbo2013}{2013} for a discussion). This
complexity makes the task of producing forecasts of future number of
births difficult.

However, forecasts of fertility are vitally important in anticipating
demand for a variety of products and services. In the short term, demand
for maternity care and nursery school places and the aggregate cost of
child benefit payments all depend on the size of birth cohorts. Over
longer horizons, future fertility is a crucial determinant of overall
population size and age structure, and consequently will affect national
accounts and fiscal sustainability (Office for Budget Responsibility
\protect\hyperlink{ref-OfficeforBudgetResponsibility2018}{2018}).

As a result, methodologies that accurately capture the degree of
uncertainty surrounding forecasts may help manage the risks surrounding
inevitable deviations from point forecasts of fertility (Bijak et al.
\protect\hyperlink{ref-Bijak2015b}{2015}). A predictive forecast
distribution can be combined with information about the costs of over-
or under-prediction and the degree of risk aversion of the decision
maker to allow a course of action to be decided upon, following the
principles of statistical decision theory (Bijak et al.
\protect\hyperlink{ref-Bijak2015b}{2015}).

In several developed countries, fertility rates in recent decades can be
observed to take on a bi-modal shape when viewed as a function of age
(Chandola, Coleman, and Hiorns
\protect\hyperlink{ref-Chandola1999}{1999}). This suggests the
possibility that the underlying population is heterogeneous with respect
to their fertility behaviour (ibid), which may reflect, for example, the
differential opportunity cost of childbearing to women of different
educational groups (Van Bavel
\protect\hyperlink{ref-VanBavel2010}{2010}; Billari and Philipov
\protect\hyperlink{ref-Billari2005}{2005}). This paper describes an
approach to the forecasting of fertility that captures this feature of
contemporary fertility patterns. The method focuses first on forecasting
the overall level of childbearing, measured by the average number of
births to mothers of particular cohorts (that is, mothers born in the
same year). Following Chandola, Coleman, and Hiorns
(\protect\hyperlink{ref-Chandola1999}{1999}), this summary measure of
fertility is subsequently decomposed over the fertile age range using a
mixture of smooth parametric curves. Extending their approach to a
forecasting context, predictions of future values are produced by
forecasting forward both the parameters of these curves and the summary
measure of the level of fertility using time series methods. Posterior
predictive distributions coherently incorporate both the underlying time
series stochasticity and uncertainty about model parameters.

The approach advocated in this paper has the advantage of parsimony,
with relatively few effective parameters needed for each cohort, and
furthermore constrains predictions of future fertility curves to the
space of shapes described by two-component mixtures. A number of authors
have found such mixtures to be useful in modelling fertility curves in
populations in the developed world (Chandola, Coleman, and Hiorns
\protect\hyperlink{ref-Chandola1999}{1999}; Peristera and Kostaki
\protect\hyperlink{ref-Peristera2007}{2007}; Bermúdez et al.
\protect\hyperlink{ref-Bermudez2012}{2012}), but this approach adds
dependence between successive cohorts within a Bayesian hierarchical
framework.

This paper makes several specific novel contributions to the literature
on fertility forecasting:

\begin{itemize}
\tightlist
\item
  The existing work on parametric mixture modelling of fertility is
  extended to a forecasting context, and a strategy for identifying the
  parameters of the two components is set out.
\item
  The efficacy of employing different combinations of parametric forms
  within the mixture model is examined.
\item
  Existing works that forecasting parametric fertility models without
  mixtures (e.g. Congdon (\protect\hyperlink{ref-Congdon1990}{1990});
  Congdon (\protect\hyperlink{ref-Congdon1993}{1993}); Knudsen, McNown,
  and Rogers (\protect\hyperlink{ref-Knudsen1993}{1993}); Mazzuco and
  Scarpa (\protect\hyperlink{ref-Mazzuco2015}{2015})) employ a two stage
  approach whereby parametric curves are first fitted to demographic
  rates, and time series models subsequently fitted to the parameters of
  such models. The method set out in this paper fits the parametric
  curves and time series models simultaneously within a Bayesian
  framework, accounting for uncertainty in both elements of the model in
  predictions.
\item
  Births are modeled using a negative binomial likelihood with smooth
  age-specific over-dispersion, allowing for a better fit across the
  most fertile ages.
\end{itemize}

Existing work on mixture forecasting in the statistical literature (e.g.
Wong and Li \protect\hyperlink{ref-Wong2000}{2000}; Li and Wong
\protect\hyperlink{ref-Li2001}{2001}; Wood, Rosen, and Kohn
\protect\hyperlink{ref-Wood2011a}{2011}) differs subtly to the approach
adopted in this paper. Firstly, the cited works apply to univariate time
series, whereas in this application the data are multivariate; at each
time point we have a set of correlated observations. Secondly, this
paper does not consider that observations themselves are drawn from a
mixture distribution, so that the likelihood for each point is a
weighted sum of density functions, but that particular parameters in a
hierarchical specification are weighted sums of parametric functions.

The rest of this paper proceeds by first describing the data used for
the forecasting application in the remainder of the paper (Section
\ref{data-and-notation}). Next, in Section
\ref{modelling-and-forecasting-fertility}, existing approaches to the
forecasting of fertility are introduced, and areas in which further
developments may be fruitful are identified. Section
\ref{model-description} presents the model used throughout the paper,
while Section \ref{estimation-and-results} describes the model
estimation process, and provides illustrative results of the application
of the model to data from England \& Wales. Section
\ref{assessing-model-variants} assesses the performance of mixtures of
different components using data from a selection of developed country.
The final section (Section \ref{discussion}) provides a discussion of
the overall contribution of the paper and directions for further work.

\section{Data and notation}\label{data-and-notation}

This section describes the data used in this applications and sets out
some of the framework and notation used in the rest of the paper. Many
of the details will be familiar to demographers, but is included for
completeness.

In most applications, the target variable for fertility forecasting is a
vector of age-specific fertility rates. Observed data on the number of
births \(b\) is classified by age of the new-born's mother at last
birthday \(x\) and the year of observation \(t\). The age-specific
fertility rate for a given calendar year can be defined in terms of
expectation of the associated random variable \(B\).

\begin{align}
\begin{split}
B_{xt} \sim \text{Poisson}(\lambda_{xt}) \\
E[B_{x,t}] = \lambda_{x,t}\\
f_{xt} = \frac{E[B_{xt}]} {R_{xt}},
\end{split}
\end{align}

where \(R_{x,t}\) is the cumulative population exposure to risk at age
\(x\) in year \(t\), measured in person-years. This latter quantity is
often approximated by the size of the female population aged \(x\) at
the midpoint of year \(t\).

However, the cohort of the mother (that is, the year in which they were
born) is considered to be particularly important in the analysis of
fertility, because it is believed that decision to bear a child is
influenced by past life experiences and by the number of children a
woman already has (Ryder \protect\hyperlink{ref-Ryder1964}{1964}). This
suggests that we might also be interested in an alternative
specification of the fertility rate \(f_{xc}\), with \(c\) indicating
cohort.

\begin{figure}
\centering
\includegraphics{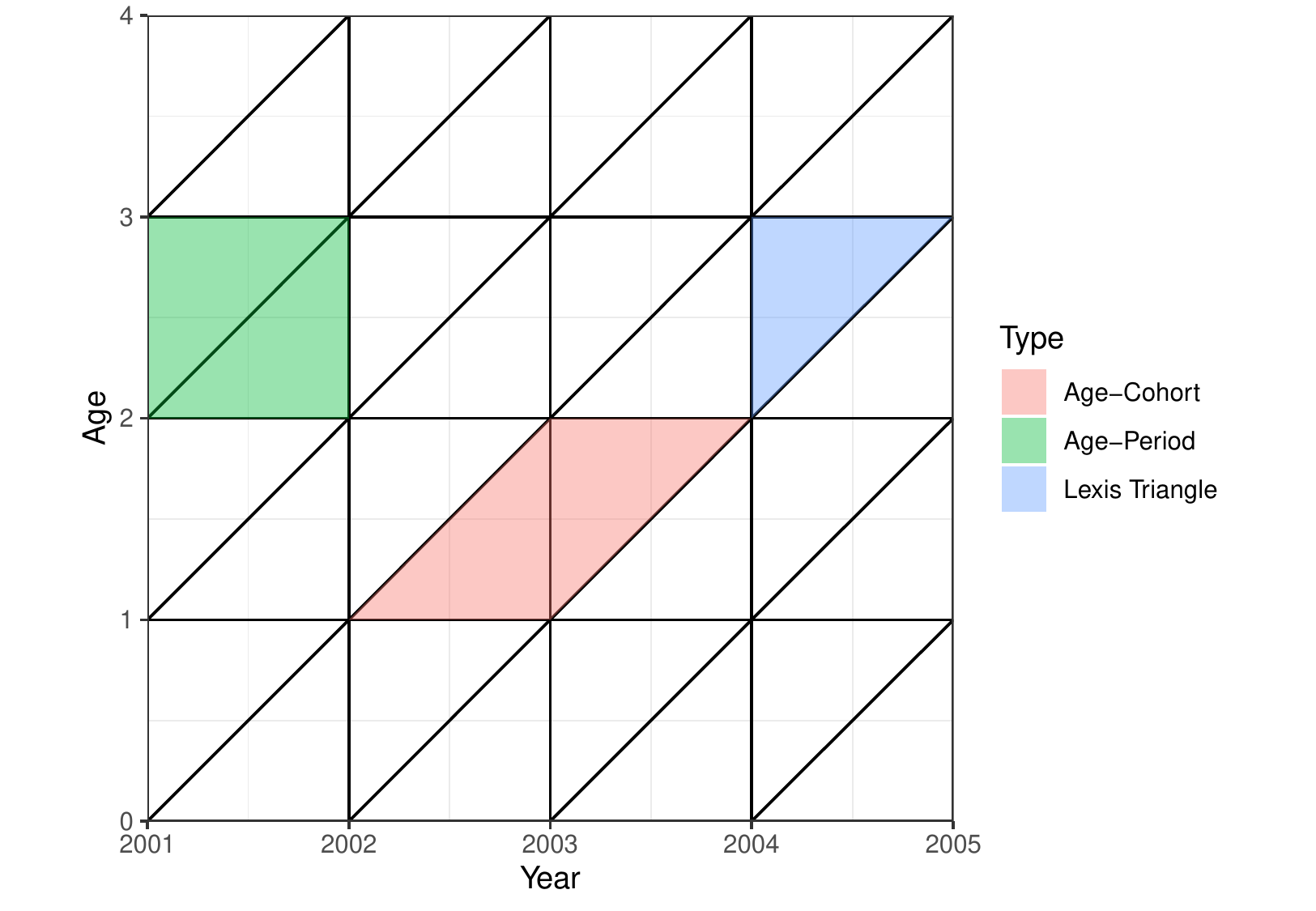}
\caption{\label{fig:lexis}Lexis diagram illustrating cells corresponding to
Age-Period, Age-Cohort, and Age-Period-Cohort classified data}
\end{figure}

When we wish to examine changes by cohort, data classified by year and
age is not ideal, as those recorded as aged \(x\) at the midpoint of
year \(t\) may have been born in year \(t-x\) or in year \(t - x - 1\).
For this work, data on births and exposure were obtained from the Human
Fertility Database (HFD) (Human Fertility Database
\protect\hyperlink{ref-HFD2018}{2018}). The HFD decomposes both birth
and exposure age-period data into approximate age-period-cohort cells
(Jasilioniene et al. \protect\hyperlink{ref-Jasilioniene2015}{2015}),
known as Lexis triangles because of the shape they make on the `Lexis
diagram', a common tool for analysing demographic data (e.g. Minton et
al. \protect\hyperlink{ref-Minton2017}{2017}). This decomposition allows
data classified by cohort and age or by cohort and period to be
reconstituted from these lower-level building blocks (Jasilioniene et
al. \protect\hyperlink{ref-Jasilioniene2015}{2015}). Figure
\ref{fig:lexis} is a Lexis diagram illustrating these differing
classifications. The life-course of an individual can be represented on
this diagram as a line which begins at the point along the horizontal
axis corresponding to their birth date, and is projected at an angle of
45 degrees clockwise from the vertical. The points at which the line
crosses vertical and horizontal grid-lines represent transitions between
calendar years and birthdays respectively. The red square covers
demographic events happening during 2001 to the population aged 2 at
their last birthday, and so corresponds to Age-Period data. In contrast,
the green parallelogram corresponds to events happening to those born in
during 2001 between the ages of 1 and 2 (Age-Cohort data). Finally, the
blue triangle is the `Lexis Triangle' corresponding to the same cohort
for events occurring while they were aged 2 during 2004
(Age-Period-Cohort data).

Age-Cohort data will be used for fitting models of fertility and
producing forecasts throughout the remainder of the paper. It is worth
noting that at a particular point in time, data for more recent cohorts
(represented as diagonal slices through the Lexis diagram) are
incomplete, in that we only have observations for younger ages. The
forecasting problem is therefore to both complete the fertility schedule
of existing cohorts and to predict the outcomes for new cohorts. It
should be said that by using the HFD Age-Cohort data as provided without
accounting for the way it has been constructed from Age-Period data,
uncertainty in the final forecasts will be underestimated. However, it
is expected that this error will be small relative to other sources of
error.

\begin{figure}
\centering
\includegraphics{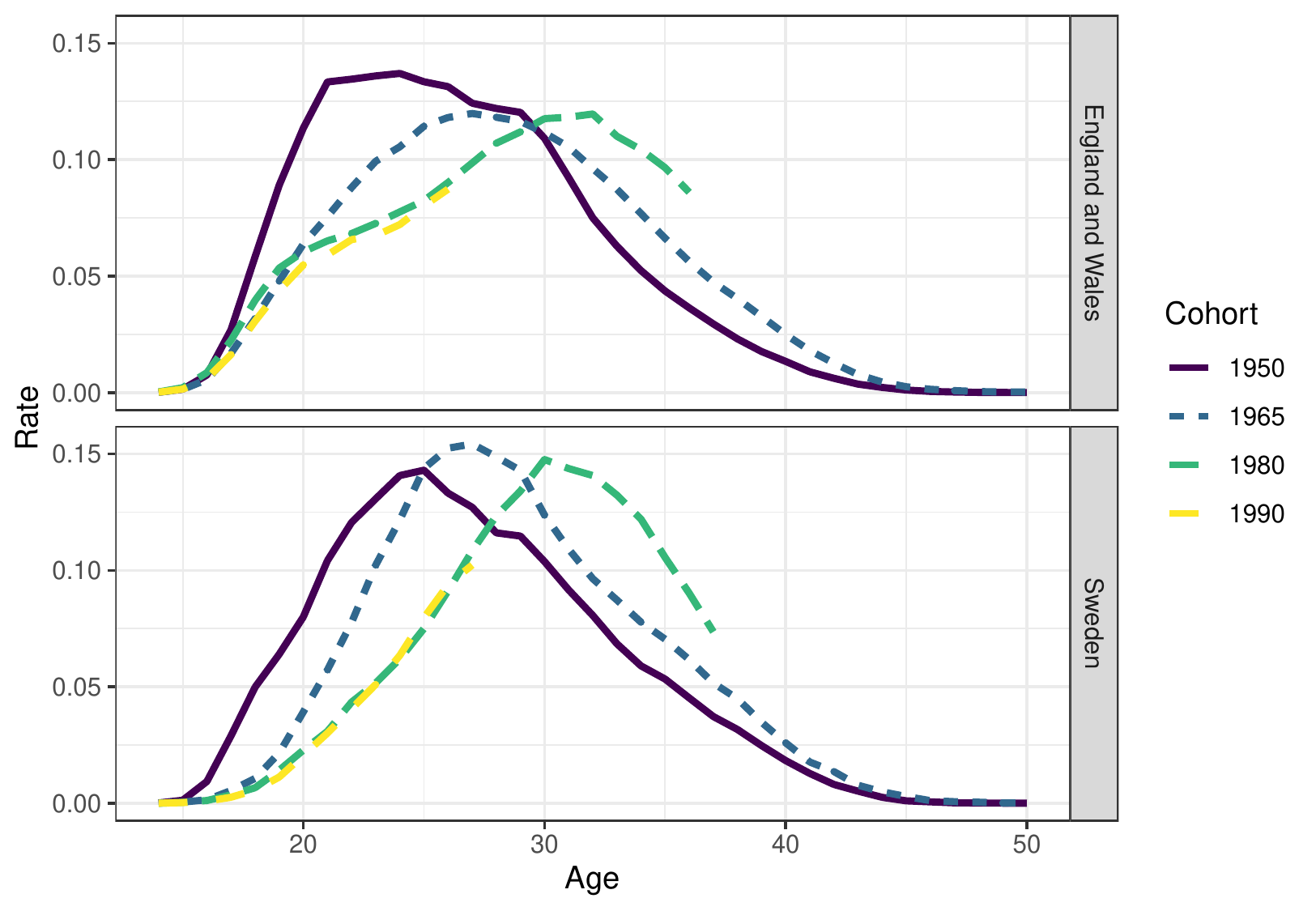}
\caption{\label{fig:empplots}Cohort fertility schedules for England and
Wales and for Sweden, selected cohorts.}
\end{figure}

Figure \ref{fig:empplots} shows fertility age profiles for selected
cohorts for Sweden and for England \& Wales (with fertility aggregated
for the latter two nations). In both countries, the curve shifts to the
right for later cohorts as women postpone childbearing as the result of
longer periods in education or greater professional prospects (Billari
et al. \protect\hyperlink{ref-Billari2007a}{2007}). However, a notable
difference between the fertility curves for later cohorts for Sweden and
England \& Wales is that in the latter case we see a distinct hump or
shoulder for younger ages, as identified by Chandola, Coleman, and
Hiorns (\protect\hyperlink{ref-Chandola1999}{1999}). Such a feature is
also visible in Irish and to a lesser extent USA data (ibid). Fertility
projection models need to be flexible enough to capture this bulge at
younger ages in order to be able to adequately forecast future fertility
rates for these countries.

It is common to describe fertility for a given calendar year or cohort
in terms of summary measures. For cohort fertility, the most natural
summary is the area under the fertility hazard function for that cohort,
which is the average number of children born to a member of this birth
cohort over the course of their lifetime, assuming they survive to the
highest possible age of childbearing \(X\) (Jasilioniene et al.
\protect\hyperlink{ref-Jasilioniene2015}{2015}). This is generally
termed Completed Family Size (CFS), and can be approximated by the sum
of the individual age-specific rates:

\begin{align}
\begin{split}
\theta_c &= \int_{12}^{50} \frac{\lambda_{ac}}{R_{ac}}da \\
&\approx \sum_{x=12}^{49}f_{x,c}
\end{split}
\end{align}

The approximation is exact under the assumption the hazard and exposure
are constant within each age interval. The equivalent period-based
measure is the well-known Total Fertility Rate (TFR), which is
constructed in the same way from Age-Period data, substituting the \(c\)
cohort index for a \(t\). This measure approximates the average number
of children born to a `synthetic' cohort who experienced a given
period's age-specific fertility rates across their whole fertile
lifespan.

\section{Modelling and Forecasting
Fertility}\label{modelling-and-forecasting-fertility}

A wide range of approaches have been deployed for modelling and
forecasting fertility, of which Booth
(\protect\hyperlink{ref-Booth2006}{2006}) and Bohk-Ewald, Li, and
Myrskylä
(\protect\hyperlink{ref-Bohk-Ewald2018a}{2018}\protect\hyperlink{ref-Bohk-Ewald2018a}{a})
provide comprehensive reviews. An instructive area of the literature
focuses on merely modelling fertility rates rather than forecasting
them. One approach which has particularly inspired the current work was
introduced by Hoem et al. (\protect\hyperlink{ref-Hoem1981}{1981}), who
used a range of density functions to approximate the shape of the
fertility curve for a given period. Density functions are useful in such
a context as they are parsimonious representations of uni-modal curves,
and are guaranteed to integrate to one. A minimum of two parameters are
needed to describe the location \(\mu\) and scale \(\tau\) of the curve,
and a separate parameter \(\theta\) representing the TFR is employed
multiplicatively to determine the area under the curve and thus the
overall level of fertility, so that

\begin{align}
f_{x} = \theta \; g(x \; ; \; \mu,\tau)
\end{align}

where \(g\) is a density function.

Chandola, Coleman, and Hiorns
(\protect\hyperlink{ref-Chandola1999}{1999}) adapted this work to
account for the changes in observed data in the 1990s. In particular,
the emergence of bi-modal fertility curves in England and the United
States necessitated the use of two parametric components in order to
obtain a reasonable approximation of the data. Peristera and Kostaki
(\protect\hyperlink{ref-Peristera2007}{2007}) and Bermúdez et al.
(\protect\hyperlink{ref-Bermudez2012}{2012}) went further, including new
families of density functions. Peristera and Kostaki
(\protect\hyperlink{ref-Peristera2007}{2007}) used split-normal
distributions where the variance parameters take different values either
side of the mean, while Bermúdez et al.
(\protect\hyperlink{ref-Bermudez2012}{2012}) employed Weibull density
functions, which may be parameterised to have either positive or
negative skews.

Parametric approaches have also been employed for forecasting fertility,
as opposed to merely modelling its age structure, although these have
tended to involve only single-component models rather than mixtures. The
double-exponential nuptial model of Coale and McNeil
(\protect\hyperlink{ref-Coale1972}{1972}) was reparameterised and
adapted by Knudsen, McNown, and Rogers
(\protect\hyperlink{ref-Knudsen1993}{1993}) to allow for forecasting
using ARIMA models, while Congdon
(\protect\hyperlink{ref-Congdon1990}{1990}) similarly employed the
function of Hadwiger (\protect\hyperlink{ref-Hadwiger1940}{1940}) in the
forecasting of fertility for London boroughs. More recently, De Iaco and
Maggio (\protect\hyperlink{ref-DeIaco2016}{2016}) used ARIMA methods to
forecast forward parameters of a gamma function fitted to Italian
fertility, and furthermore used a Markov field model to capture
correlations in the error structure of this model. In keeping with the
approach adopted in this paper, Mazzuco and Scarpa
(\protect\hyperlink{ref-Mazzuco2015}{2015}) attempt to capture and
forecast the bimodal structure of fertility using a Flexible
Generalisable Skew-Normal Distribution, although the parameters in this
model are hard to interpret demographically.

The debate in the demographic literature around whether fertility is
better understood as a cohort (Ryder
\protect\hyperlink{ref-Ryder1964}{1964}) or period (Ní Bhrolcháin
\protect\hyperlink{ref-Bhrolchain1992}{1992}) phenomenon has strongly
influenced the forecasting literature. The argument revolves around the
observation that period fertility fluctuates in response to economic or
political circumstances in particular years, but often the Completed
Family Size (or Children Ever Born, CEB) of women in different cohort
remains fairly stable. This is because women adjust the timing of their
planned births (referred to as the tempo) in response to circumstances,
but the overall level (or quantum) is less subject to adjustment
(Bongaarts and Feeney \protect\hyperlink{ref-Bongaarts1998}{1998}).

The greater stability of cohort fertility has led to a number of
approaches that focus on `completing' cohort fertility. This involves
inferring the fertility rates for cohort which have already started
childbearing. For example, Evans
(\protect\hyperlink{ref-Evans1986}{1986}) completes cohort fertility
based by predicting fertility for ages 25-45 using regressions that use
the level and approximate gradient with respect to age of fertility
rates between 15-25 as covariates. Similarly, Congdon
(\protect\hyperlink{ref-Congdon1993}{1993}) adapts his previous work
(Congdon \protect\hyperlink{ref-Congdon1990}{1990}) by using Hadwiger
functions to complete cohort fertility schedules from age 30 onwards,
and subsequently forecasts forward Hadwiger parameters using time series
methods. De Beer (\protect\hyperlink{ref-DeBeer1985}{1985}) fits ARIMA
models to smooth fertility rates in both the cohort and age direction,
so that information from both previous cohorts and from earlier ages can
be used to complete cohort fertility schedules. Cheng and Lin
(\protect\hyperlink{ref-Cheng2010}{2010}) use a more traditional
Age-Period-Cohort model to incorporate effects along both time-varying
dimensions.

Myrskylä, Goldstein, and Cheng
(\protect\hyperlink{ref-Myrskyla2013a}{2013}) demonstrate a pragmatic
method for completing cohort fertility whereby for each age, the average
change in the fertility rate over the preceding 5-year period is assumed
to persist for 5 years into the future. The authors describe this as the
`freeze-slope' approach, for obvious reasons, and this is contrasted
with naive `freeze-rate' approach, in which rates are held constant.
Schmertmann et al. (\protect\hyperlink{ref-Schmertmann2014}{2014})
incorporate elements of this approach within a conjugate normal-normal
Bayesian model that generates prior distributions that smooth fertility
rates in the both the age and cohort direction. The approach extracts
three principal components from a large collection of historical data
taken from the Human Fertility Database. One element of the constructed
prior penalises deviations from linear combinations of these principal
components. An additional element of the prior penalises deviations from
both `freeze-rate' and `freeze-slope' projections. The priors are
calibrated so that the size of the penalty reflects the distribution of
such deviations in historical data. This approach is developed by
Ellison, Forster, and Dodd (\protect\hyperlink{ref-Ellison2018}{2018}),
who construct a hierarchical Bayesian model maintaining the underlying
assumptions of the model of Schmertmann et al.
(\protect\hyperlink{ref-Schmertmann2014}{2014}).

An alternative approach focuses more on extrapolation. Lee
(\protect\hyperlink{ref-Lee1993}{1993}) adapts the principle
component-based method of Lee and Carter
(\protect\hyperlink{ref-leecarter1992}{1992}), originally developed for
mortality forecasting, to forecast fertility along the period axis. Li
and Wu (\protect\hyperlink{ref-Li2003}{2003}) extend this method using
cohort as the primary axis, while Hyndman and Ullah
(\protect\hyperlink{ref-Hyndman2007}{2007}) provide a more general
approach within the functional data framework. Wiśniowski et al.
(\protect\hyperlink{ref-Wisniowski2015}{2015}) further adapt the Lee and
Carter (\protect\hyperlink{ref-leecarter1992}{1992}) approach as part of
a comprehensive Bayesian population forecasting framework.

The work of Ševčíková et al.
(\protect\hyperlink{ref-Sevcikova2015}{2015}) is of particular interest
because it has been employed by the United Nations for the World
Population Prospects since 2015. This method uses a Bayesian
hierarchical model to describe the global distribution of schedules of
TFR evolution, noting that countries tend to follow a common pattern of
initially high fertility, decline, and later stabilisation. Expert
opinion is utilised to determine target patterns of fertility over age
(the so-called Proportionate Age Specific Rates) for each country.

A recent comprehensive assessment of methods for completing cohort
fertility schedules established that few perform better than the naive
`freeze rate' technique (Bohk-Ewald, Li, and Myrskylä
\protect\hyperlink{ref-Bohk-Ewald2018a}{2018}\protect\hyperlink{ref-Bohk-Ewald2018a}{a}).
Comparing the projected CFS against observed data for each method over a
wide range of countries and forecast jump-off years, the authors found
that only the methods developed by Myrskylä, Goldstein, and Cheng
(\protect\hyperlink{ref-Myrskyla2013a}{2013}), Schmertmann et al.
(\protect\hyperlink{ref-Schmertmann2014}{2014}), Ševčíková et al.
(\protect\hyperlink{ref-Sevcikova2015}{2015}), and De Beer
(\protect\hyperlink{ref-DeBeer1985}{1985}) consistently outperformed the
naive forecast. However, errors at the level of age-specific fertility
rates were not assessed, and performance in forecasting \emph{new}
cohorts (which is essential for population predictions) was also not
tested.

The strategy adopted in this paper builds on the parametric approaches
of Hoem et al. (\protect\hyperlink{ref-Hoem1981}{1981}), Peristera and
Kostaki (\protect\hyperlink{ref-Peristera2007}{2007}) and Bermúdez et
al. (\protect\hyperlink{ref-Bermudez2012}{2012}), fitting explicit time
series models for the parameters defining the model for each year or
cohort. Although the work of Bohk-Ewald, Li, and Myrskylä
(\protect\hyperlink{ref-Bohk-Ewald2018a}{2018}\protect\hyperlink{ref-Bohk-Ewald2018a}{a})
identified poor performance on the part of the parametric models of Hoem
et al. (\protect\hyperlink{ref-Hoem1981}{1981}) and Peristera and
Kostaki (\protect\hyperlink{ref-Peristera2007}{2007}), the original
models do not involve any dependence between adjacent cohorts or
calendar years. This work rectifies this short--coming, while retaining
the advantages of these models, namely the parsimony of the parametric
approach, the ability to separate the overall level of fertility and the
shape of the curve, and the restriction of the shape of the fertility
curve to a family of plausible values.

\section{Model Description}\label{model-description}

This section describes in detail the different elements of the model,
and the justifications behind them. Starting with the definition of the
likelihood, births at each age are assumed to be distributed according
to a negative binomial distribution (parameterised according to the
mean), so that

\begin{align}
B_{xc} \sim \text{Negative Binomial}(R_{xc}f_{xc}, \text{exp}(\phi_x)),
\end{align}

where \(B_{xc}\) are the number of children born to women of cohort
\(c\) between exact ages \(x\) and \(x+1\), \(R_{xc}\) is the exposure
to risk between these ages for this cohort, \(f_{xc}\) is the
corresponding age-specific cohort fertility rate, and \(\phi_x\) is a
parameter controlling the degree of over-dispersion relative to the
Poisson distribution, which varies with age \(x\).

The fertility rate \(f_{xc}\) is modelled as product of \(\theta_c\), a
parameter describing the average number of births to cohort \(c\) by
maximum childbearing age \(X\), and \(\xi_{xc}\), where the vector of
\(\pmb{\xi}_{c}\) parameters for each time period describes the age
distribution of the fertility rate schedule (termed the Proportionate
Age-Specific Fertility Rate (PASFR) by Ševčíková et al.
(\protect\hyperlink{ref-Sevcikova2015}{2015})), so that:

\begin{align}
\begin{split}
f_{xc} = \; \theta_c \xi_{xc}, \;
\sum_{x=14}^{X}\xi_{xc} = \; 1 \;, \text{for all} \; c 
\end{split}
\end{align}

The age distribution \(\pmb{\xi_{c}}\) is modelled as a mixture of two
components, each parameterised by a location and a spread parameter. In
particular, as with Hoem et al.
(\protect\hyperlink{ref-Hoem1981}{1981}), probability density functions
are used for the components as these have the convenient property that
they integrate to one. Alternatively, other functions with this property
could also be used. For a particular age \(x\), an approximation of the
mass of the distribution between \(x\) and \(x+1\), weighted by the
mixture parameter \(\psi\), defines that component's contribution to
\(\xi\), so that

\begin{align}
\begin{split}
\label{eq:model}
\tau_{x,c} =& \psi_c \;g_1(x + \tfrac{1}{2}; \; \mu^{(1)}_{c}, \tau^{(1)}_{c}) + (1-\psi_c) \; g_2(x + \tfrac{1}{2}; \; \mu^{(2)}_{c}, \tau^{(2)}_{c}) \\
\psi \in&  \;[0,1]
\end{split}
\end{align}

where \(g_i(x ; \mu,\tau)\) is the value at \(x\) of the density
function with location \(\mu_c^{(i)}\) and spread \(\tau_c^{(i)}\), and
\(\psi_c\) is the mixture parameter. The superscripts \((i)\) on the
location and spread parameters distinguish between the two mixture
components.

\subsection{Modelling evolution over
time}\label{modelling-evolution-over-time}

The specification in Equation \eqref{eq:model} leaves us with six unique
parameters for each time period: two location parameters, two scale
parameters, a mixture-weight parameter, and a parameter describing the
overall level of fertility. One practical problem is ensuring that the
two components are distinct to avoid identifiability and label switching
issues. One strategy to circumvent this is to force the location of one
component be strictly greater than the other. This can be achieved by
re-parameterisation to use the sum and difference of the locations
rather than the raw values, combined with enforcing constraints on these
new parameters so that they are positive and lie within reasonable
bounds. Given the range of fertile ages (\textasciitilde{}14-50 year),
the sum of the locations should lie above \(35\), and to allow
identification of the two mixture components, the gap is constrained to
be at least \(2\) years.

After this re-parameterisation, the resultant set of parameters
\(\pmb{\eta}_c\) can then by forecast forwards using auto-regressive
models,

\begin{align}
\begin{split}
\pmb{\eta}_c &= \{ \theta_c,\; \psi_c,\; \mu^{(s)}_c, \mu^{(d)}_c, \tau^{(1)}_c, \tau^{(2)}_c \} \\
\pmb{\eta}_c &\sim  \text{N}\left(A\pmb{\eta}_{c-1}, \Sigma_{\eta} \right), \\
\end{split}
\end{align}

where superscripts \(s\) and \(d\) indicate the sum and difference of
the location parameters respectively. A special case of this class of
models is the simple independent random walk:

\begin{align}
\begin{split}
\pmb{\eta}_c = \pmb{\eta}_{c - 1} + \pmb{\epsilon}_c \\
\pmb{\epsilon}_c \sim N(\pmb{0}, \pmb{\sigma^{T}_{\eta}}I).
\end{split}
\end{align}

Priors for the first elements of all the parameters modelled as time
series are chosen to be weakly informative. Flexibility as to the exact
forecasting model is possible; more complicated model classes, including
ARIMA models incorporating differencing and moving average elements, or
stochastic volatility models, could also be considered.

\subsection{Model Components}\label{model-components}

Building on the models discussed previously by Hoem et al.
(\protect\hyperlink{ref-Hoem1981}{1981}), Chandola, Coleman, and Hiorns
(\protect\hyperlink{ref-Chandola1999}{1999}), and Bermúdez et al.
(\protect\hyperlink{ref-Bermudez2012}{2012}), we examine three possible
parametric functions for the mixture components; the gamma density, the
Hadwiger or inverse Gaussian density, and the Weibull density. We wish
to adopt non-standard parameterisation for each of these functions so
that they are defined in terms of a location and a spread parameter. In
this way, elements of the model that are to be given time series priors
have a meaningful interpretation that is relatively consistent
regardless of the specific function used. For the gamma component, the
mode and standard deviation are used, while for the Hadwiger density the
mean is instead adopted as the location parameter. For the Weibull
density, the form of the density function necessitates other choices;
the median and the distance from the median to the upper quartile are
instead used for the location and spread parameters. The split normal
distribution of Peristera and Kostaki
(\protect\hyperlink{ref-Peristera2007}{2007}) is not used, as although
the authors find it fits the data relatively well, the join in the
function at the mode makes it conceptually unappealling.

Although Chandola, Coleman, and Hiorns
(\protect\hyperlink{ref-Chandola1999}{1999}), Peristera and Kostaki
(\protect\hyperlink{ref-Peristera2007}{2007}) and Bermúdez et al.
(\protect\hyperlink{ref-Bermudez2012}{2012}) use the same density
function for both the components (\(g_1\) and \(g_2\)) of their
respective parametric mixtures, there are advantages to combining
different parametric functions within a mixture, as the different
functions used have different tail behaviour, and may be better suited
to approximating fertility at older or younger ages. For example, the
Weibull density function can skew in either direction, allowing a
steeper decline than is possible with the positively-skewed gamma and
Hadwiger distributions. These distinctions may help provide for better
identification of the two components. As a result, we examined all
possible combinations of the three functions.

\subsection{Overdispersion}\label{overdispersion}

The Poisson distribution, a natural choice for modelling count data, has
variance equal to its mean. This assumption can be too restrictive in
many cases, and the use of the negative binomial distribution allows for
over-dispersion relative to the Poisson, which tends to lead to smoother
estimates of rates. However, the effect of over-dispersion is more
pronounced in absolute terms for larger counts, and so the use of a
single over-dispersion parameter for fertility can lead to a greater
emphasis on the fit in the tails of the age-specific rate distribution
where counts are small than in the more fertile stretches of the age
range. To address this concern, we specify a smooth, age-specific
over-dispersion function using a penalised basis-spline (Wood
\protect\hyperlink{ref-Wood2006}{2006}), and allow the data to determine
where greater over-dispersion is needed. A random walk prior penalises
deviations from smoothness in the basis function coefficients (Lang and
Brezger \protect\hyperlink{ref-Lang2001}{2001}):

\begin{align}
\begin{split}
\phi_x &= \pmb{\beta} \pmb{B}(x)\\
\beta_i &\sim \text{Normal}(\beta_{i-1}, \; \sigma_{\phi}^{2}).
\end{split}
\label{eq:disp} 
\end{align}

where \(\pmb{B}(x)\) is a matrix of B-spline basis functions.

Fertility data is often aggregated over open age intervals at the
beginning and end of the fertile age range. For example, in the UK,
births to women of ages 49 and above are reported in aggregate. In
contrast to the Poisson case, the sum of \(n\) negative binomial
distributions with different means does not follow a standard
distribution, so for simplicity the distribution of births in these
aggregate groups is approximated by a negative binomial distribution
with the appropriate mean and variance. Only a small proportion of
births occur to women in these age groups, so the approximation is
unlikely to make a significant difference to results.

\section{Estimation and Results}\label{estimation-and-results}

The model has been tested on data from the Human Fertility Database for
four countries: USA; England \& Wales (in combination); Sweden; and
France. Each of these countries have reasonably long time series
available, although there are distinct differences in the patterns
evident in the data. The USA and England \& Wales display bimodality to
a greater degree than France and Sweden. Although the Human Fertility
Database provides birth data from ages 12-55, these data are
extrapolated at the youngest and oldest ages where empirical information
is not available (Jasilioniene et al.
\protect\hyperlink{ref-Jasilioniene2015}{2015}). As a result, HFD data
on births below age 15 and above 49 are aggregated for the purposes of
model fitting.

The latest 10 years of data were held back for each country and not used
in fitting, to provide for an out-of-sample assessment of the model
described above. The No-U-Turn Sampler (NUTS) variant of Hamiltonian
Monte Carlo (Hoffman and Gelman
\protect\hyperlink{ref-Hoffman2014}{2014}), available in the
\texttt{Stan} software package, was used to obtained posterior samples
of the model (Stan Development Team
\protect\hyperlink{ref-StanDevelopmentTeam2018}{2018}). Hamiltonian
Monte Carlo simulates movement through the parameter space by analogy to
a physical system where the potential energy is equal to negative
log-posterior (Neal \protect\hyperlink{ref-Neal2010}{2010}).

Four chains, each consisting of 10000 iterations, were run for each
combination of model specification and country. Nine different models
(involving all combinations of the three parametric functions) were
tested for each of the four countries, giving a total of 36 different
sets of posterior samples. The first half of each chain was used as a
warm-up period during which the sampling parameters were allowed to
adapt to the shape of the posterior. With respect to the adaptation
parameters, a slightly higher target acceptance rate (0.95) and maximum
NUTS tree-depth (12) was used than is selected by default, providing for
smaller integrator step-sizes and longer NUTS trajectories (Stan
Development Team \protect\hyperlink{ref-StanDevelopmentTeam2019}{2019}).
The remaining samples, thinned by a factor of 4 to prevent excessive
memory use, were used for posterior inference.

For the majority of model and country combinations, Gelman-Rubin split
\(\hat{r}\) diagnostics are below the suggested 1.05 threshold and
examination of trace-plots indicate sampler convergence to the target
distribution (Gelman and Rubin \protect\hyperlink{ref-Gelman1992}{1992};
Stan Development Team
\protect\hyperlink{ref-StanDevelopmentTeam2019}{2019}). In a minority of
cases (7 out of 36), diagnostics indicated problems were encountered
during sampling. In particular, sampling of models involving a Hadwiger
distribution was found to be difficult, particularly when the Hadwiger
was used to model the component corresponding to older ages.

In one case -- the Weibull/Hadwiger model applied to French data -- the
two parametric components were not distinguishable and obtaining
posterior samples was not possible. Multi-modal posteriors were obtained
in 4 cases, with one or more chains sampling from a mode located in a
different area of the parameter space to the others, generally because
the mixture parameters approach a limit for some cohorts with incomplete
data. In these cases, either the mode occupied by the majority of
chains, or, where the four chains are divided equally between two modes,
the mode with the smallest Leave-One-Out Information Criterion (LOOIC)
(Vehtari, Gelman, and Gabry \protect\hyperlink{ref-Vehtari2016}{2016})
was chosen for posterior inference, and the other chains discarded.
Recomputed values of the split \(\hat{r}\) statistic indicate that in
each case the remaining chains appear to have converged to the selected
mode. By choosing a mode based on LOOIC for this small subset of models
means that where posteriors are multi-modal, only the mode that is best
able to predict left-out data-points is considered in the model
comparison exercises that follow in Section
\ref{assessing-model-variants}.

Divergent transitions were encountered in two additional cases.
Divergent transitions indicate that the value of Hamiltonian in the
No-U-Turn sampler differs greatly over the course of a HMC trajectory
due to integration errors (Stan Development Team
\protect\hyperlink{ref-StanDevelopmentTeam2019}{2019}). In practice this
means that the curvature of the posterior varies such that a single
leap-frog integrator step-size is insufficient to explore all areas of
the posterior. In both cases, diagnostic plots show that this problem
was caused by significant posterior probability being placed on low
values of one of the time series innovation variance parameters, which
results in a narrower posterior in this area of the parameter space than
elsewhere. The sampler is unable to reach these areas of the posterior,
meaning the obtained posterior samples are truncated away from zero for
this parameter (Betancourt
\protect\hyperlink{ref-Betancourt2016}{2016}). Re-parameterisation to a
non-centred model where standard normal innovations are sampled
separately and multiplied by the variance parameter may help resolve
this problem (ibid), but also makes enforcing constraints on the time
series more difficult. Results from these models are not reported, but
qualitative examination of the truncated posteriors suggest it is
unlikely that these parameter combinations would be among the best
performing models.

The code used to fit the model and generate the results presented in
this section are available from github: \url{https://github.com/jasonhilton/fert_mix}

\subsection{Illustrative Results for England \&
Wales}\label{illustrative-results-for-england-wales}

\begin{figure}
\centering
\includegraphics{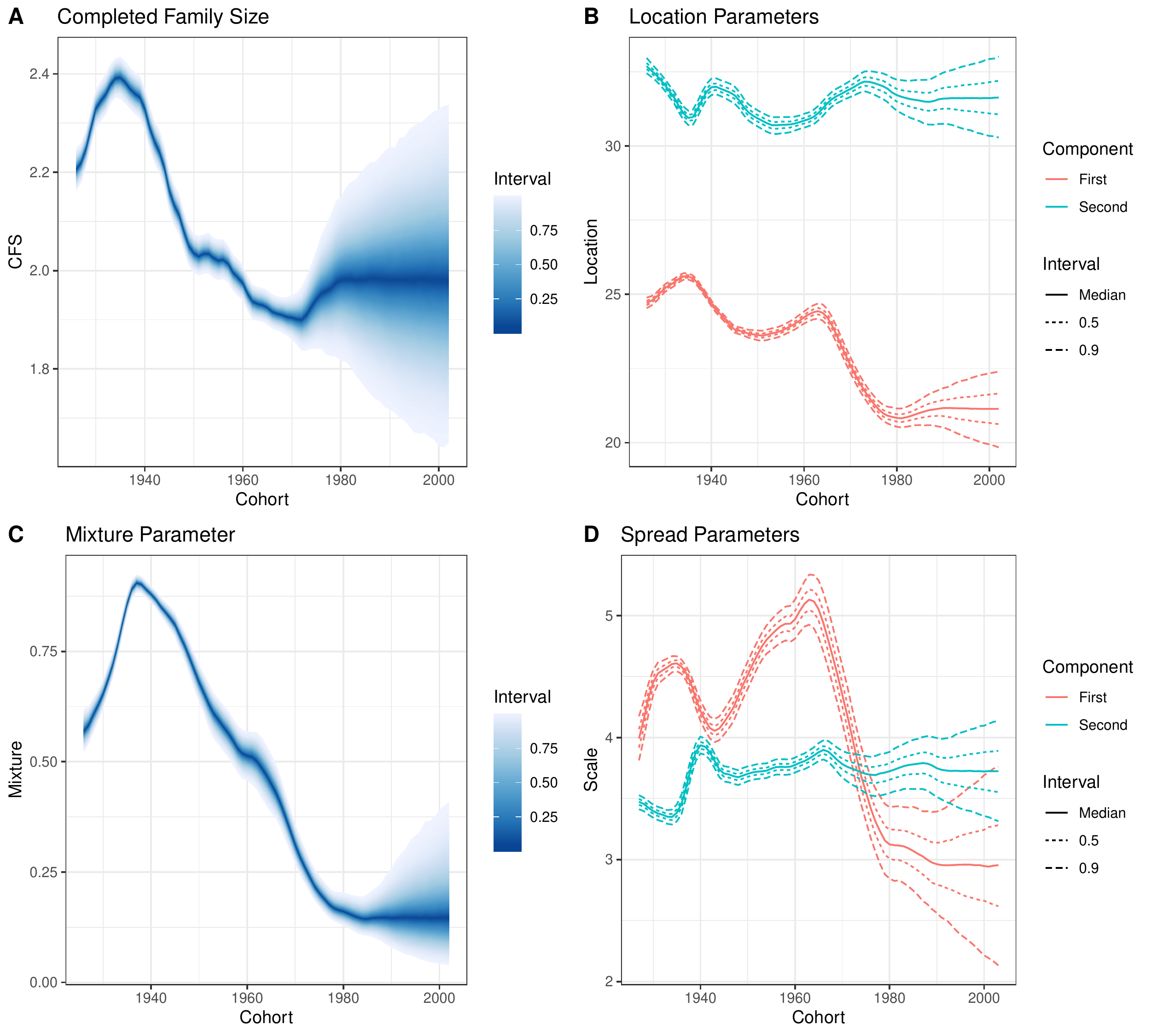}
\caption{\label{fig:paramENW}Posterior Distribution of Model Parameters;
gamma/Weibull Mixture; England and Wales data}
\end{figure}

We first examine an example model fitted with data from the England \&
Wales. In this case, we employ a gamma density for the parametric
component corresponding primarily to younger ages, and similarly a
Weibull density for older ages. The over-dispersion in births is allowed
to vary by age in accordance with Equation \eqref{eq:disp}. All
time-varying parameters are given random walk priors. The posterior
distributions are displayed in Figure \ref{fig:paramENW}. Birth data is
fully observed for all ages up until the 1966 cohort, who were 50 in the
last year of observation and thus deemed to have completed childbearing.
The posterior distributions for incomplete cohorts is informed by both
their partially complete childbearing experiences and the time series
priors on the model parameters, which penalise excessive deviations from
past values. This combination predicts a slight increase in completed
family size in the short term, before leveling off at around 2 children
per woman.

The mode of the lower gamma component fluctuates between ages 20 and 25,
while the median of the upper Weibull density is considerably higher, at
around 33 years (Figure \ref{fig:paramENW}B). The weight given to the
younger component decreases after the parents of the baby-boom
generation have completed their childbearing, as individuals begin to
delay fertility to later ages (Figure \ref{fig:paramENW}C). Uncertainty
in all parameters increases for later cohorts for which there are fewer
or no observations.

\begin{figure}
\centering
\includegraphics{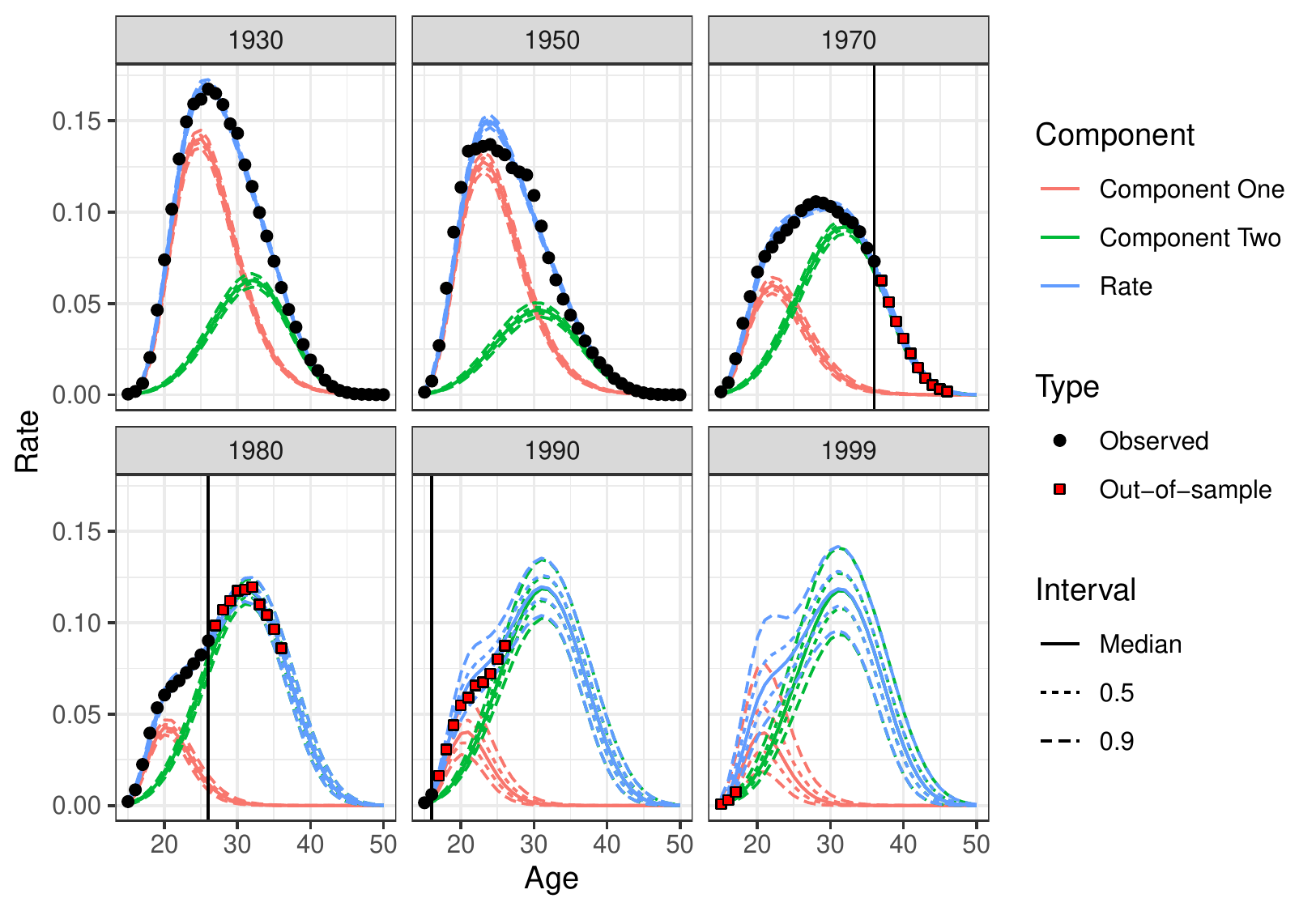}
\caption{\label{fig:ewrates}Posterior Distribution of Age-Specific Fertility
Rates for Selected Cohort. Vertical black lines for later cohorts
represent forecast jump-off points.}
\end{figure}

Figure \ref{fig:ewrates} presents posterior distributions for the two
weighted components and their sum, the mean Age-Specific Fertility Rate,
for six selected cohorts. This plot shows how the size, position and
weight of the two component changes over cohorts. A slight lack of fit
is observed for the 1950 cohort, which is caused by the inability of the
model to entirely capture the period fluctuations associated with the
baby bust. Qualitatively, it also shows the model performs reasonably
well at predicting the out-of-sample observations, with most
observations close to the posterior median, although a more formal
assessment taking into account the negative binomial uncertainty
associated with prediction of births is required to have confidence in
its performance. Similarly, the choices regarding the specification of
functional components should be justified in terms of predictive
performance. The next section examines the forecasting error and
coverage of different variants of the model to this end.

\section{Assessing Model Variants}\label{assessing-model-variants}

In order to choose between the different possible specifications of our
model, we wish to make a formal assessment of the performance of the
model both in predicting observations falling within the fitting period
and in forecasting new observations. To this end, for each possible
configuration of functional components (for example, gamma-Weibull or
gamma-gamma) the model was estimated on data up to 2006 for four
countries separately.

The approximate Leave-One-Out (LOO) cross-validation approach introduced
for model assessment by Vehtari, Gelman, and Gabry
(\protect\hyperlink{ref-Vehtari2016}{2016}) and (implemented in the
\texttt{loo} R package) is used to assess the ability of the various
model specifications to predict data drawn from same data-generating
process. This method uses importance sampling to produce an
approximation to the expected log point-wise predictive density (ELPD)
obtained by predicting every point in the data-set using only the rest
of the data, thus avoiding the need to refit the model \(N\) times
leaving out each data point. Pareto smoothing is used reduce the
instability in the tails of the importance sampling weights (Vehtari,
Gelman, and Gabry (\protect\hyperlink{ref-Vehtari2016}{2016}), termed
Pareto Smoothed Importance Sampling, or PSIS).

For some models, a small number of observations exhibited somewhat high
values of the k parameter of the Pareto distribution fitted to the
importance weights. This indicates the PSIS estimates of that points
contribution to the ELPD are unreliable, likely due to overly
influential observations (Vehtari, Gelman, and Gabry
\protect\hyperlink{ref-Vehtari2016}{2016}). In these cases, the model
was re-estimated a maximum of three times with one or more unreliable
observations left out, allowing the contributions of these points to the
ELPD value to be calculated directly and combined with the existing PSIS
estimates to provide the final ELPD value. Where estimates for more than
3 data-points were unreliable, several points were left out in each
re-fitting to save on computation, with these points grouped to maximise
the distance between them within each refitting. In practice, this
process of refitting made little difference to the overall ELPD scores
or the model comparisons.

\begin{figure}
\centering
\includegraphics{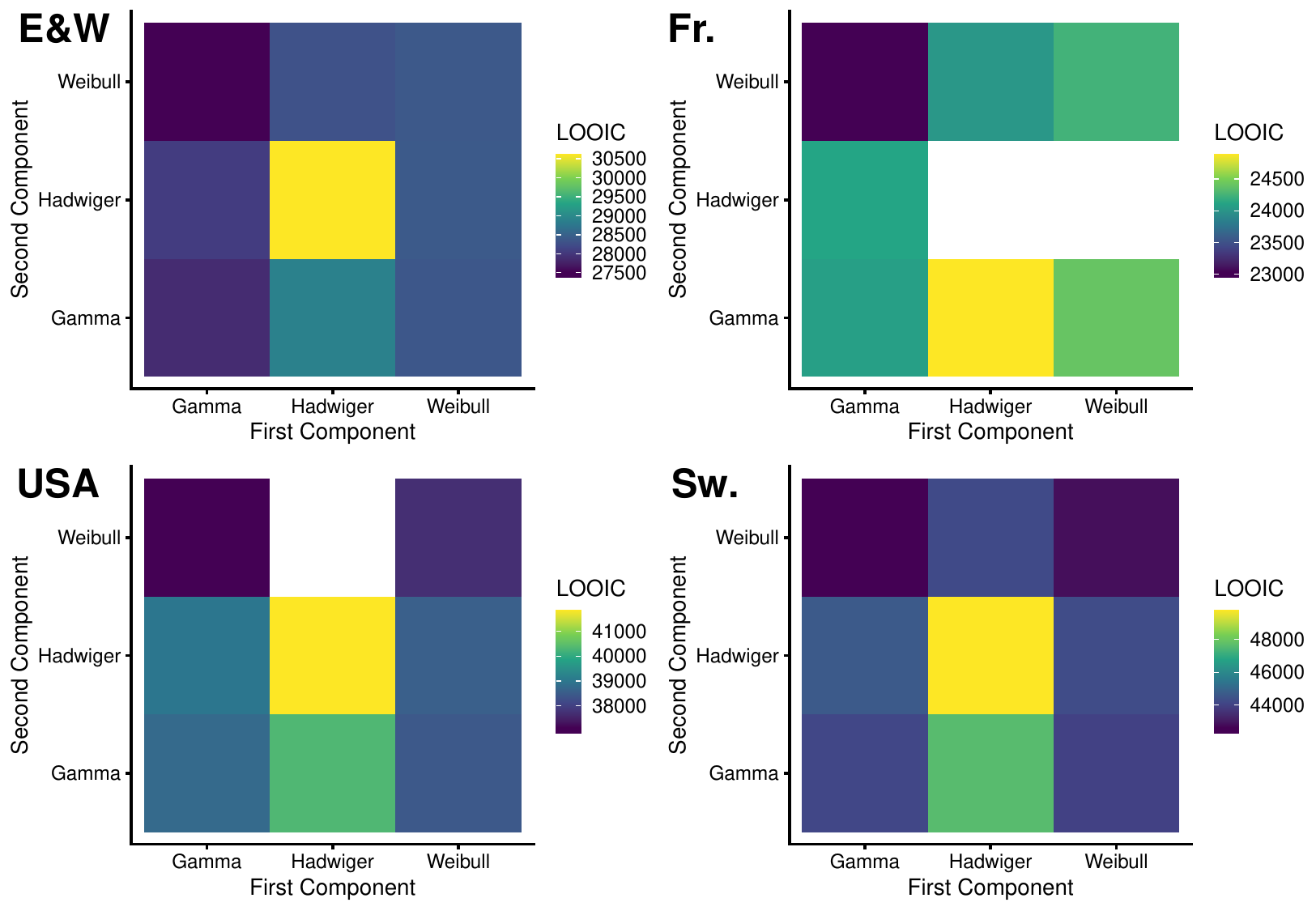}
\caption{\label{fig:LOOIC}Pareto Smoothed Importance Sampling appoximation
to the Leave-One-Out Information Criterion for all combinations of
Parametric Components for England and Wales, France, USA, and Sweden}
\end{figure}

Figure \ref{fig:LOOIC} displays results in terms of the Leave-One-Out
Information Criterion, which transforms the ELPD so that it is on the
same scale as the deviance and other commonly used model assessment
measures, such as the Deviance Information Criterion and the Bayesian
Information Criterion. The four sub-plots each represent one of the four
countries used in the assessment exercise, while within each subplot,
the first parametric component is represented in the columns and the
second components are in the rows. Thus, the top-left cell refers to a
model with a gamma component for younger ages and a Weibull Component
for older ages. Missing cells correspond to models for which posterior
samples are not available, as detailed above. The deeper blue the cell
of the plot, the better the model is at predicting left-out points
within the range of the data. For all countries, the gamma-Weibull model
obtains the best LOOIC value, although given the sampling variation of
the LOOIC metric, it is not always possible to definitively separate it
from the next best performing models. The Weibull-Weibull and
gamma-gamma model also perform relatively well across countries,
although in the French case, the gamma-Weibull model was the clear
winner.

Although specifying models that are able to capture the properties of
the observed data well enough to predict left-out points is important,
this paper is primarily focused on forecasting. Forecasts of fertility
rates are compared to actual observed forecasts for the years 2007-2016
(As the data used is indexed by cohort, the held-back sample is not
strictly speaking for specific years, but for an index t = c + a, which
will consist of data from two calendar years). The root mean squared
error (RMSE) and the empirical coverage over all age-specific rates is
calculated for each model to assess both the adequacy of the point
forecasts and the calibration of the forecast uncertainty. With respect
to coverage, for each forecast we calculate the proportion of
corresponding observations of age-specific rates that fall within the
specified probabilistic predictive internals (90\% and 50\% in this
instance). Because we are observing only one set of correlated outcomes
for each country (age-specific rates for 2007-2016), rather than many
independent repetitions, this metric does not correspond exactly to the
frequentist concept of coverage, but it does however give an indication
as to whether the uncertainty in forecasts appear well calibrated
compared to actual observed deviations from predictions.

Table \ref{tab:rmsetable} shows a comparison for each model of the RMSE
of the mean posterior forecast over the 10 years of hold-back data
2007-2016. As well as variants of the model set out in this paper,
results are also reported for the naive forecast, and for models
identified by Bohk-Ewald, Li, and Myrskylä
(\protect\hyperlink{ref-Bohk-Ewald2018}{2018}\protect\hyperlink{ref-Bohk-Ewald2018}{b})
as performing well in predicting cohort fertility. The best performing
models for each country are in bold, while an NA indicates that
obtaining results for that combination of components was not possible.
It is clear that parametric models are competitive with the comparison
models. The Weibull-Weibull model, for instance, has lower forecast
error than both the Myrskylä, Goldstein, and Cheng
(\protect\hyperlink{ref-Myrskyla2013a}{2013}) and Schmertmann et al.
(\protect\hyperlink{ref-Schmertmann2014}{2014}) models across all four
countries, although it is less successful than the freeze rate model for
USA and Sweden. The gamma-Weibull model, which was consistently amongst
the models with the strongest predictive power for points covered by the
fitting period, similarly outperformed all competitors in forecasting
England \& Wales and USA data.

\begin{table}[t]

\caption{\label{tab:rmsetable}Comparison of the 10-year out-of sample forecast Root Mean Squared Error over different countries and models}
\centering
\begin{tabular}{lllll}
\toprule
\multicolumn{1}{c}{Model} & \multicolumn{4}{c}{RMSE} \\
\cmidrule(l{3pt}r{3pt}){1-1} \cmidrule(l{3pt}r{3pt}){2-5}
Model & E and W & France & Sweden & USA\\
\midrule
\addlinespace[0.3em]
\multicolumn{5}{l}{\textbf{Parametric Models}}\\
\hspace{1em}Gamma / Gamma & 0.0069 & 0.0049 & 0.0069 & \textbf{0.0090}\\
\hspace{1em}Gamma / Hadwiger & 0.0083 & 0.0044 & 0.0068 & 0.0091\\
\hspace{1em}Gamma / Weibull & 0.0053 & 0.0071 & 0.0073 & 0.0092\\
\hspace{1em}Hadwiger / Gamma & 0.0142 & 0.0158 & 0.0233 & 0.0167\\
\hspace{1em}Hadwiger / Hadwiger & 0.0119 & NA & 0.0180 & 0.0160\\
\hspace{1em}Hadwiger / Weibull & 0.0059 & 0.0096 & 0.0089 & NA\\
\hspace{1em}Weibull / Gamma & 0.0075 & \textbf{0.0038} & 0.0048 & 0.0114\\
\hspace{1em}Weibull / Hadwiger & 0.0076 & NA & 0.0113 & 0.0110\\
\hspace{1em}Weibull / Weibull & \textbf{0.0052} & 0.0041 & 0.0067 & 0.0111\\
\addlinespace[0.3em]
\multicolumn{5}{l}{\textbf{Other Models}}\\
\hspace{1em}Schmertmann et al. 2014 & 0.0073 & 0.0066 & 0.0082 & 0.0159\\
\hspace{1em}Myrskyl{\"a} et al. 2013 & 0.0067 & 0.0063 & 0.0101 & 0.0121\\
\hspace{1em}Freeze Rates & 0.0068 & 0.0051 & \textbf{0.0040} & 0.0119\\
\bottomrule
\end{tabular}
\end{table}

Tables \ref{tab:coverage50} and \ref{tab:coverage90} set out the the
proportion of held-out observations that fall within the 50\% and 90\%
predictive intervals respectively. As was also clear from the point
forecast errors, US data appears relatively difficult to forecast, with
considerable under-coverage at the 50\% interval for all models. In
contrast, for UK data, coverage for most models falls reasonably close
to the ideal values. Coverage properties of models with Weibull or gamma
as the first component appear reasonably good, in contrast to those
models which use Hadwiger density functions for younger ages, which tend
to under-cover. The Schmertmann et al.
(\protect\hyperlink{ref-Schmertmann2014}{2014}) model also has strong
coverage properties, perhaps because this model is calibrated based on
historical data across multiple countries, while the model of Myrskylä,
Goldstein, and Cheng (\protect\hyperlink{ref-Myrskyla2013a}{2013})
appears to systematically undercover, in agreement with the results in
Bohk-Ewald, Li, and Myrskylä
(\protect\hyperlink{ref-Bohk-Ewald2018}{2018}\protect\hyperlink{ref-Bohk-Ewald2018}{b}).

\begin{table}[t]

\caption{\label{tab:coverage50}Comparison of the 10-year out-of-sample 50\% empirical forecast coverage over different countries and models }
\centering
\begin{tabular}{llllll}
\toprule
 & Model & E and W & France & Sweden & USA\\
\midrule
\addlinespace[0.3em]
\multicolumn{6}{l}{\textbf{Parametric Models}}\\
\hspace{1em} & Gamma / Gamma & 0.57 & 0.40 & 0.46 & 0.22\\

\hspace{1em} & Gamma / Hadwiger & 0.52 & 0.44 & 0.48 & 0.22\\

\hspace{1em} & Gamma / Weibull & 0.67 & 0.30 & 0.54 & 0.24\\

\hspace{1em} & Hadwiger / Gamma & 0.20 & 0.24 & 0.24 & 0.19\\

\hspace{1em} & Hadwiger / Hadwiger & 0.32 & NA & 0.32 & 0.19\\

\hspace{1em} & Hadwiger / Weibull & 0.50 & 0.48 & 0.62 & NA\\

\hspace{1em} & Weibull / Gamma & 0.56 & 0.61 & 0.63 & 0.37\\

\hspace{1em} & Weibull / Hadwiger & 0.52 & NA & 0.51 & 0.43\\

\hspace{1em} & Weibull / Weibull & 0.57 & 0.64 & 0.50 & 0.34\\
\cmidrule{1-6}
\addlinespace[0.3em]
\multicolumn{6}{l}{\textbf{Other Models}}\\
\hspace{1em} & Schmertmann et al. 2014 & 0.49 & 0.60 & 0.60 & 0.36\\

\hspace{1em} & Myrskyl{\"a} et al. 2013 & 0.18 & 0.08 & 0.11 & 0.11\\
\bottomrule
\end{tabular}
\end{table}

\begin{table}[t]

\caption{\label{tab:coverage90}Comparison of the 10-year out-of sample forecast empirical 90\% coverage over different countries and models }
\centering
\begin{tabular}{llllll}
\toprule
 & Model & E and W & France & Sweden & USA\\
\midrule
\addlinespace[0.3em]
\multicolumn{6}{l}{\textbf{Parametric Models}}\\
\hspace{1em} & Gamma / Gamma & 0.80 & 0.81 & 0.77 & 0.77\\

\hspace{1em} & Gamma / Hadwiger & 0.80 & 0.90 & 0.77 & 0.79\\

\hspace{1em} & Gamma / Weibull & 0.95 & 0.71 & 0.92 & 0.49\\

\hspace{1em} & Hadwiger / Gamma & 0.58 & 0.62 & 0.61 & 0.52\\

\hspace{1em} & Hadwiger / Hadwiger & 0.82 & NA & 0.63 & 0.76\\

\hspace{1em} & Hadwiger / Weibull & 0.84 & 0.75 & 0.92 & NA\\

\hspace{1em} & Weibull / Gamma & 0.86 & 0.99 & 0.90 & 0.79\\

\hspace{1em} & Weibull / Hadwiger & 0.83 & NA & 0.64 & 0.79\\

\hspace{1em} & Weibull / Weibull & 0.82 & 0.99 & 0.82 & 0.84\\
\cmidrule{1-6}
\addlinespace[0.3em]
\multicolumn{6}{l}{\textbf{Other Models}}\\
\hspace{1em} & Schmertmann et al. 2014 & 0.79 & 0.76 & 0.72 & 0.62\\

\hspace{1em} & Myrskyl{\"a} et al. 2013 & 0.41 & 0.28 & 0.23 & 0.25\\
\bottomrule
\end{tabular}
\end{table}

Interestingly, the model is able to capture fertility trends in Sweden
and France despite these countries displaying considerably more
uni-modal fertility schedules than in the USA or England \& Wales. The
ability of the proposed family of models to capture differing shapes and
trends in the tails of the schedule may explain its utility in these
cases.

Overall, for the countries studied it is not easy to state that one
parametric model specification is definitively better than another; the
models best able to predict left-out data from the period to 2006 did
not necessarily have the better RMSE or coverages over the period
2007-2016. However, the gamma-Weibull model appears to offer a
reasonable compromise, as it appears to consistently produce good
results across contexts.

\section{Discussion}\label{discussion}

This paper has introduced a flexible family of models for the
forecasting of parametric mixtures, designed with the particular
application of forecasting bi-modal fertility in mind. The model can
capture the shape of both bi- and uni-modal age-specific fertility
curve, and projects forward the level of fertility together with
parameters relating to the location, scale and mixture weight of the two
parametric components. The assessment of the proposed family of models
across a small range of countries show that they produce reasonable
forecasts of recent fertility trends that are competitive with models
identified in the literature as being best able to predict cohort
fertility (Bohk-Ewald, Li, and Myrskylä
\protect\hyperlink{ref-Bohk-Ewald2018}{2018}\protect\hyperlink{ref-Bohk-Ewald2018}{b}).

A relatively simplistic variant of the model has been proposed here, but
there is considerable potential for expansion, particularly in the area
of the time series models used for forecasting; fully-fledged ARIMA
models will enable the sort of immediate continuation of short-term
trends that is a feature of the models of Myrskylä, Goldstein, and Cheng
(\protect\hyperlink{ref-Myrskyla2013a}{2013}) and Schmertmann et al.
(\protect\hyperlink{ref-Schmertmann2014}{2014}). Capturing correlations
between various model parameters may also enhance the predictive
capacities of the model, to the extent that such correlations persist
over time.

In order to more fully evaluate the wider applicability of the model to
the forecasting of fertility, a more comprehensive assessment exercise
should be carried out along the lines of Bohk-Ewald, Li, and Myrskylä
(\protect\hyperlink{ref-Bohk-Ewald2018a}{2018}\protect\hyperlink{ref-Bohk-Ewald2018a}{a}),
whereby forecasts errors are evaluated over a large range of countries
and across many time periods. The present discussion has focused on the
ability of the model to predict cohort trends in fertility, but
forecasting along the period dimension is equally possible (e.g. Congdon
\protect\hyperlink{ref-Congdon1993}{1993}). Looking further afield, the
generality of the model described means that it may also be appropriate
in other contexts in which bi-modal curves must be forecast into the
future.

\section*{Acknowledgement}\label{acknowledgement}
\addcontentsline{toc}{section}{Acknowledgement}

This work was supported by the ESRC Centre for Population Change - phase
II (grant ES/K007394/1) at the University of Southampton. The use of the
IRIDIS High Performance Computing Facility, and associated support
services at the University of Southampton, in the completion of this
work is also acknowledged. All the views presented in this paper are
those of the authors only.

\section*{References}\label{references}
\addcontentsline{toc}{section}{References}

\hypertarget{refs}{}
\hypertarget{ref-Balbo2013}{}
Balbo, Nicoletta, Francesco C Billari, and Melinda Mills. 2013.
``Fertility in Advanced Societies: A Review of Research: La fécondité
dans les sociétés avancées: un examen des recherches.'' \emph{European
Journal of Population = Revue Europeenne de Demographie} 29 (1): 1--38.
doi:\href{https://doi.org/10.1007/s10680-012-9277-y}{10.1007/s10680-012-9277-y}.

\hypertarget{ref-Bermudez2012}{}
Bermúdez, S., R. Blanquero, J.A. Hernández, and J. Planelles. 2012. ``A
new parametric model for fitting fertility curves A new parametric model
for fitting fertility curves.'' \emph{Population Studies} 66 (3):
297--310.

\hypertarget{ref-Betancourt2016}{}
Betancourt, Michael. 2016. ``Diagnosing Suboptimal Cotangent
Disintegrations in Hamiltonian Monte Carlo.''

\hypertarget{ref-Bijak2015b}{}
Bijak, Jakub, Isabel Alberts, Juha Alho, John Bryant, Thomas Buettner,
Jane Falkingham, Jonathan J. Forster, et al. 2015. ``Letter to the
Editor: Probabilistic population forecasts for informed decision
making.'' \emph{Journal of Official Statistics} 31 (4): 537--44.
doi:\href{https://doi.org/10.1037/emo0000122.Do}{10.1037/emo0000122.Do}.

\hypertarget{ref-Billari2005}{}
Billari, Francesco C., and Dimiter Philipov. 2005. ``Women's education
and entry into a first union. A simultaneous-hazard comparative analysis
of Central and Eastern Europe.'' \emph{Vienna Yearbook of Population
Research} 1 (2004): 91--110.
doi:\href{https://doi.org/10.1553/populationyearbook2004s91}{10.1553/populationyearbook2004s91}.

\hypertarget{ref-Billari2007a}{}
Billari, Francesco C., Hans Peter Kohler, Gunnar Andersson, and Hans
Lundström. 2007. ``Approaching the limit: Long-term trends in late and
very late fertility.'' \emph{Population and Development Review} 33 (1):
149--70.
doi:\href{https://doi.org/10.1111/j.1728-4457.2007.00162.x}{10.1111/j.1728-4457.2007.00162.x}.

\hypertarget{ref-Bohk-Ewald2018a}{}
Bohk-Ewald, Christina, Peng Li, and Mikko Myrskylä. 2018a. ``Forecast
accuracy hardly improves with method complexity when completing cohort
fertility.'' \emph{Proceedings of the National Academy of Sciences} 115
(3): 630--81.
doi:\href{https://doi.org/10.1073/pnas.1722364115}{10.1073/pnas.1722364115}.

\hypertarget{ref-Bohk-Ewald2018}{}
---------. 2018b. ``Supplemental Material for 'Forecast accuracy hardly
improves with method complexity when completing cohort fertility'.''
\emph{Proceedings of the National Academy of Sciences} 115 (3): 630--81.
\url{www.pnas.org/cgi/doi/10.1073/pnas.1722364115}.

\hypertarget{ref-Bongaarts1998}{}
Bongaarts, John, and Griffith Feeney. 1998. ``On the Quantum and Tempo
of Fertility.'' \emph{Population and Development Review} 24 (2):
271--91.

\hypertarget{ref-Booth2006}{}
Booth, Heather. 2006. ``Demographic Forecasting: 1980 to 2005 in
review.'' \emph{Interational Journal of Forecasting} 22: 547--81.
\url{http://www.sciencedirect.com/science/article/pii/S016920700600046X}.

\hypertarget{ref-Chandola1999}{}
Chandola, T, D A Coleman, and R W Hiorns. 1999. ``Recent European
fertility patterns: fitting curves to 'distorted' distributions.''
\emph{Population Studies} 53 (3): 317--29.
doi:\href{https://doi.org/10.1080/00324720308089}{10.1080/00324720308089}.

\hypertarget{ref-Cheng2010}{}
Cheng, Roger, and Eric S. Lin. 2010. ``Completing incomplete cohort
fertility schedules.'' \emph{Demographic Research} 23: 223--56.
doi:\href{https://doi.org/10.4054/DemRes.2010.23.9}{10.4054/DemRes.2010.23.9}.

\hypertarget{ref-Coale1972}{}
Coale, A. J., and D. R. McNeil. 1972. ``The distribution by age of the
frequency of first marriage in a female cohort.'' \emph{Journal of the
American Statistical Association} 67 (340): 743--49.
doi:\href{https://doi.org/10.1080/01621459.1972.10481287}{10.1080/01621459.1972.10481287}.

\hypertarget{ref-Congdon1990}{}
Congdon, Peter. 1990. ``Graduation of Fertility Schedules: An Analysis
of Fertility Patterns in London in the 1980s and an Application to
Fertility Forecasts.'' \emph{Regional Studies} 24 (4): 311--26.
doi:\href{https://doi.org/10.1080/00343409012331346014}{10.1080/00343409012331346014}.

\hypertarget{ref-Congdon1993}{}
---------. 1993. ``Statistical Graduation in Local Demographic Analysis
and Projection.'' \emph{Journal of the Royal Statistical Society (Series
A)} 156 (2): 237--70.

\hypertarget{ref-DeBeer1985}{}
De Beer, J. 1985. ``A time series model for cohort data.'' \emph{J Am
Stat Assoc} 80 (391): 525--30.
\url{http://www.ncbi.nlm.nih.gov/entrez/query.fcgi?cmd=Retrieve\&db=PubMed\&dopt=Citation\&list_uids=12267306}.

\hypertarget{ref-DeIaco2016}{}
De Iaco, Sandra, and Sabrina Maggio. 2016. ``A dynamic model for
age-specific fertility rates in Italy.'' \emph{Spatial Statistics} 17.
Elsevier Ltd: 105--20.
doi:\href{https://doi.org/10.1016/j.spasta.2016.05.002}{10.1016/j.spasta.2016.05.002}.

\hypertarget{ref-Ellison2018}{}
Ellison, Joanne, Jonathan J Forster, and Erengul Dodd. 2018.
``Forecasting of cohort fertility under a hierarchical Bayesian
approach.''
\url{https://eprints.soton.ac.uk/cgi/eprintbypureuuid?uuid=dead073b-5d4c-47af-b4bc-13c9dfdea9e6}.

\hypertarget{ref-Evans1986}{}
Evans, M. D. R. 1986. ``American Fertility Patterns : A Comparison of
White and Nonwhite Cohorts Born 1903-56.'' \emph{Population and
Development Review} 12 (2): 269--93.

\hypertarget{ref-Gelman1992}{}
Gelman, Andrew, and Donald B Rubin. 1992. ``Inference from Iterative
Simulation Using Multiple Sequences.'' \emph{Statistical Science} 7 (4):
457--72.

\hypertarget{ref-Hadwiger1940}{}
Hadwiger, H. 1940. ``Eine analytische reproduktionsfunktion für
biologische gesamtheiten.'' \emph{Scandinavian Actuarial Journal} 3-4:
101--13.
doi:\href{https://doi.org/10.1080/03461238.1940.10404802}{10.1080/03461238.1940.10404802}.

\hypertarget{ref-Hoem1981}{}
Hoem, Jan M., Dan Madsen, Jorgen Lovgreen Nielsen, Else-Marie Ohlsen,
Hans Oluf Hansen, and Bo Rennermalm. 1981. ``Experiments in Modelling
Recent Danish Fertility Curves.'' \emph{Demography} 18 (2): 231--44.

\hypertarget{ref-Hoffman2014}{}
Hoffman, M D, and Andrew Gelman. 2014. ``The no-U-turn sampler:
Adaptively setting path lengths in Hamiltonian Monte Carlo.''
\emph{Journal of Machine Learning Research} 15 (2008): 1--31.
\url{http://arxiv.org/abs/1111.4246}.

\hypertarget{ref-HFD2018}{}
Human Fertility Database. 2018. ``Human Fertility Database.'' Max Planck
Institute for Demographic Research (Germany); Vienna Institute of
Demography (Austria). \url{www.humanfertility.org}.

\hypertarget{ref-Hyndman2007}{}
Hyndman, Rob J., and Md. Shahid Ullah. 2007. ``Robust forecasting of
mortality and fertility rates: A functional data approach.''
\emph{Computational Statistics and Data Analysis} 51 (10): 4942--56.
doi:\href{https://doi.org/10.1016/j.csda.2006.07.028}{10.1016/j.csda.2006.07.028}.

\hypertarget{ref-Jasilioniene2015}{}
Jasilioniene, A., D. A. Jdanov, T Sobotka, E. M. Andreev, K. Zeman, and
V. M. Shkolnikov. 2015. ``Methods Protocol for the Human Mortality
Database.'' Human Fertility Database.
\url{https://www.humanfertility.org/Docs/methods.pdf}.

\hypertarget{ref-Knudsen1993}{}
Knudsen, Christin, Robert McNown, and Andrei Rogers. 1993. ``Knudsen,
Christin McNown, Robert Rogers, Andrei.'' \emph{Social Science Research}
22: 1--23.

\hypertarget{ref-Lang2001}{}
Lang, Stefan, and Andreas Brezger. 2001. ``Bayesian P-Splines.'' Insitut
Für Statistik Sonderforschungsbereich 386. Munich:
Ludwig-Maximilians-Universität-München.
\url{https://epub.ub.uni-muenchen.de/1617/}.

\hypertarget{ref-Lee1993}{}
Lee, R D. 1993. ``Modeling and forecasting the time series of US
fertility: age distribution, range, and ultimate level.''
\emph{International Journal of Forecasting} 9 (2): 187--202.
doi:\href{https://doi.org/10.1016/0169-2070(93)90004-7}{10.1016/0169-2070(93)90004-7}.

\hypertarget{ref-leecarter1992}{}
Lee, Ronald D, and Lawrence R Carter. 1992. ``Modeling and Forecasting
U.S Mortality.'' \emph{Journal of the American Statistical Association}
87 (419). American Statistical Association: pp. 659--71.
\url{http://www.jstor.org/stable/2290201}.

\hypertarget{ref-Li2003}{}
Li, Nan, and Zheng Wu. 2003. ``Forecasting cohort incomplete fertility:
A method and an application.'' \emph{Population Studies} 57 (3):
303--20.
doi:\href{https://doi.org/10.1080/0032472032000137826}{10.1080/0032472032000137826}.

\hypertarget{ref-Li2001}{}
Li, W. K., and C. S. Wong. 2001. ``On a logistic mixture autoregressive
model.'' \emph{Biometrika} 88 (3): 833--46.

\hypertarget{ref-Mazzuco2015}{}
Mazzuco, Stefano, and Bruno Scarpa. 2015. ``Fitting Age-Specific
Fertility Rates By a Skew-Symmetric Probability Density Function.''
\emph{Journal of the Royal Statistical Society (Series A)} 178 (1):
187--203.

\hypertarget{ref-Minton2017}{}
Minton, Jon, Richard Shaw, Mark A. Green, Laura Vanderbloemen, Frank
Popham, and Gerry McCartney. 2017. ``Visualising and quantifying 'excess
deaths' in Scotland compared with the rest of the UK and the rest of
Western Europe.'' \emph{Journal of Epidemiology and Community Health} 71
(5): 461--67.
doi:\href{https://doi.org/10.1136/jech-2016-207379}{10.1136/jech-2016-207379}.

\hypertarget{ref-Myrskyla2013a}{}
Myrskylä, M, Joshua R Goldstein, and Yen-Hsin Alice Cheng. 2013. ``New
Cohort Fertility Forecasts for the Developed World: Rises, Falls, and
Reversals.'' \emph{Population and Development Review} 39 (1): 31--56.
\url{http://onlinelibrary.wiley.com/doi/10.1111/j.1728-4457.2013.00572.x/abstract}.

\hypertarget{ref-Neal2010}{}
Neal, Radford. 2010. ``MCMC using Hamiltonian Dynamics.'' In
\emph{Handbook of Markov Chain Monte Carlo}, edited by Steve Brooks,
Andrew Gelman, Galin Jones, and Xiao-Li Meng Meng. Chapman; Hall / CRC
Press.

\hypertarget{ref-Bhrolchain1992}{}
Ní Bhrolcháin, Maire. 1992. ``Period Paramount? A Critique of the Cohort
Approach to Fertility.'' \emph{Population and Development Review} 18
(4): 599. doi:\href{https://doi.org/10.2307/1973757}{10.2307/1973757}.

\hypertarget{ref-OfficeforBudgetResponsibility2018}{}
Office for Budget Responsibility. 2018. ``Fiscal Sustainability
Report.'' July. Office for Budget Responsibility.
doi:\href{https://doi.org/10.1142/S021812749900095X}{10.1142/S021812749900095X}.

\hypertarget{ref-Peristera2007}{}
Peristera, Paraskevi, and Anastasia Kostaki. 2007. ``Modeling fertility
in modern populations.'' \emph{Demographic Research} 16.
doi:\href{https://doi.org/10.4054/DemRes.2007.16.6}{10.4054/DemRes.2007.16.6}.

\hypertarget{ref-Ryder1964}{}
Ryder, N B. 1964. ``The Process of Demographic Translation.''
\emph{Demography} 1 (1): 74--82.

\hypertarget{ref-Schmertmann2014}{}
Schmertmann, Carl, Emilio Zagheni, Joshua R. Goldstein, and Mikko
Myrskylä. 2014. ``Bayesian Forecasting of Cohort Fertility.''
\emph{Journal of the American Statistical Association} 109 (506):
500--513.
doi:\href{https://doi.org/10.1080/01621459.2014.881738}{10.1080/01621459.2014.881738}.

\hypertarget{ref-StanDevelopmentTeam2018}{}
Stan Development Team. 2018. ``The Stan Core Library.''
\url{http://mc-stan.org}.

\hypertarget{ref-StanDevelopmentTeam2019}{}
---------. 2019. ``Stan Modeling Language Users Guide and Reference
Manual.'' \url{http://mc-stan.org/index.html}.

\hypertarget{ref-Sevcikova2015}{}
Ševčíková, Hana, Nan Li, Vladimíra Kantorová, Patrick Gerland, and
Adrian E. Raftery. 2015. ``Age-Specific Mortality and Fertility Rates
for Probabilistic Population Projections.''
doi:\href{https://doi.org/10.1007/978-3-319-26603-9_15}{10.1007/978-3-319-26603-9\_15}.

\hypertarget{ref-VanBavel2010}{}
Van Bavel, Jan. 2010. ``Choice of study discipline and the postponement
of motherhood in europe: The impact of expected earnings, gender
composition, and family attitudes.'' \emph{Demography} 47 (2): 439--58.

\hypertarget{ref-Vehtari2016}{}
Vehtari, Aki, Andrew Gelman, and Jonah Gabry. 2016. ``Practical Bayesian
model evaluation using leave-one-out cross-validation and WAIC.''
\emph{Statistics and Computing} 27 (5). Springer US: 1--20.
doi:\href{https://doi.org/10.1007/s11222-016-9696-4}{10.1007/s11222-016-9696-4}.

\hypertarget{ref-Wisniowski2015}{}
Wiśniowski, Arkadiusz, Peter W. F. Smith, Jakub Bijak, James Raymer, and
Jonathan J. Forster. 2015. ``Bayesian Population Forecasting: Extending
the Lee-Carter Method.'' \emph{Demography} 52: 1035--59.
doi:\href{https://doi.org/10.1007/s13524-015-0389-y}{10.1007/s13524-015-0389-y}.

\hypertarget{ref-Wong2000}{}
Wong, Chun Shan, and Wai Keung Li. 2000. ``On a mixture autoregressive
model.'' \emph{Journal of the Royal Statistical Society (Series B)} 62
(1): 95--115.

\hypertarget{ref-Wood2011a}{}
Wood, Sally, Ori Rosen, and Robert Kohn. 2011. ``Bayesian Mixtures of
Autoregressive Models.'' \emph{Journal of Computational and Graphical
Statistics} 20 (1): 174--95.
doi:\href{https://doi.org/10.1198/jcgs.2010.09174}{10.1198/jcgs.2010.09174}.

\hypertarget{ref-Wood2006}{}
Wood, Simon N. 2006. \emph{Generalised Additive Models: An Introduction
with R}. Boca Raton: Chapman; Hall / CRC Press.

\end{document}